\documentclass[twocolumn,showpacs,showkeys,preprintnumbers,amsmath,amssymb,prb]{revtex4}
\usepackage{graphicx}
\usepackage{dcolumn}
\usepackage{bm}

\begin{document}

\preprint{APS/123-QED}

\title{Exact solution of the geometrically frustrated spin-1/2 Ising-Heisenberg model \\
on the triangulated Kagom\'e (triangles-in-triangles) lattice}
\author{Jozef Stre\v{c}ka} 
\email{jozef.strecka@upjs.sk, jozkos@pobox.sk} 
\homepage{http://158.197.33.91/~strecka}
\author{Lucia \v{C}anov\'a}
\author{Michal Ja\v{s}\v{c}ur}
\affiliation{Department of Theoretical Physics and Astrophysics, 
Faculty of Science, \\ P. J. \v{S}af\'{a}rik University, Park Angelinum 9,
040 01 Ko\v{s}ice, Slovak Republic}
\author{Masayuki Hagiwara}
\affiliation{KYOKUGEN (Center for Quantum Science and Technology under Extreme Conditions), 
Osaka University, 1-3 Machikaneyama, Toyonaka, Osaka 560-8531, Japan}

\date{\today}
             
\begin{abstract}
The geometric frustration of the spin-1/2 Ising-Heisenberg model on the triangulated Kagom\'e (triangles-in-triangles) lattice is investigated within the framework of an exact analytical method 
based on the generalized star-triangle mapping transformation. Ground-state and finite-temperature 
phase diagrams are obtained along with other exact results for the partition function, Helmholtz free energy, internal energy, entropy, and specific heat, by establishing a precise mapping relationship to the corresponding spin-1/2 Ising model on the Kagom\'e lattice. It is shown that the residual entropy 
of the disordered spin liquid phase is for the quantum Ising-Heisenberg model significantly lower 
than for its semi-classical Ising limit ($S_0/N_{\rm T} k_{\rm B} = 0.2806$ and $0.4752$, respectively), which implies that quantum fluctuations partially lift a macroscopic degeneracy 
of the ground-state manifold in the frustrated regime. The investigated model system has an obvious relevance to a series of polymeric coordination compounds Cu$_9$X$_2$(cpa)$_6$ (X=F, Cl, Br and cpa=carboxypentonic acid) for which we made a theoretical prediction about the temperature 
dependence of zero-field specific heat. 
\end{abstract}


\maketitle

\section{\label{sec:intro}Introduction}

The antiferromagnetic quantum Heisenberg model (AF-QHM) defined on \textit{geometrically frustrated 
planar lattices} represents a long-standing theoretical challenge due to a rich variety of unusual 
ground states it exhibits as a result of the mutual interplay between quantum fluctuations and geometric frustration.\cite{Mis04,Ric04,Hon04,LhuMis} In particular, the extensive theoretical 
studies of the spin-1/2 AF-QHM on the triangular,\cite{triang} Kagom\'e,\cite{kagome1,kagome2} Shastry-Sutherland,\cite{ssl} star,\cite{star} checkerboard,\cite{check} square Kagom\'e lattice\cite{sqkag} and others,\cite{others} have revealed a great diversity in their ground-state 
and low-temperature behavior. It is now widely accepted that the ground state of the spin-1/2 AF-QHM 
on some geometrically frustrated planar lattices like a triangular lattice is the N\'eel-like ordered state,\cite{triang} while there is a still controversial debate whether \cite{kagome1} or not \cite{kagome2} the disordered spin liquid state is the true ground state of this model 
on Kagom\'e lattice. Anyway, the macroscopic degeneracy of the ground-state 
manifold turns out to be strongly related to a geometric topology of the underlying lattice and 
it is therefore of particular research interest to explore a connection between the zero-point 
entropy and the lattice geometry. 

Another striking feature, which currently attracts a great deal of attention to the spin-1/2 
AF-QHM on the geometrically frustrated planar lattices, is being a presence of quantized 
magnetization plateaux in their low-temperature magnetization curves.\cite{Ric04,Hon04} 
It is worthwhile to remark that this outstanding quantum phenomenon has already been 
experimentally observed in several prototypical examples of the frustrated quantum spin systems 
such as the triangular lattice compounds CsFe(SO$_4$)$_2$,\cite{Ina96} Cs$_2$CuBr$_4$,\cite{Ono03} 
and RbFe(MoO$_4$)$_2$,\cite{Ina96,Pro03} the Kagom\'e lattice compound [Cu$_3$(titmb)$_2$(CH$_3$COO)$_6$].H$_2$O,\cite{Nar04} and the Shastry-Sutherland lattice compounds SrCu$_2$(BO$_3$)$_2$\cite{Kag99} and RB$_4$ (R = Er, Tm).\cite{Mic06} It also has been demonstrated 
that the interplay between the geometric frustration and quantum fluctuations might be a driving 
force for an enhanced magnetocaloric effect emerging during the adiabatic demagnetization.\cite{Zhi03} 
This makes from geometrically frustrated spin systems especially promising refrigerant materials 
in view of reaching temperatures in a sub-Kelvin range, since they often remain disordered down to 
the lowest achievable temperatures unlike paramagnetic salts usually exhibiting a spin-glass transition.  
 
\begin{figure*}
\begin{center}
\includegraphics[width=13cm]{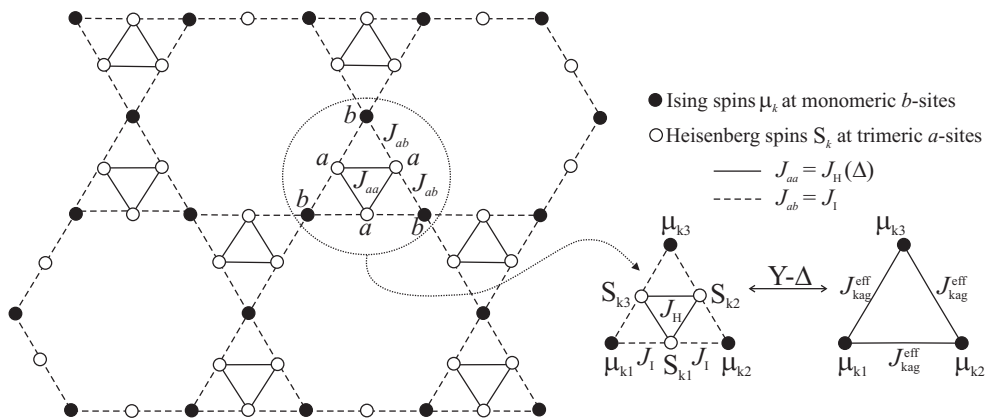} 
\end{center} 
\vspace{-4mm} 
\caption{The cross-section of the triangulated Kagom\'e (triangles-in-triangles) lattice.
The empty and full circles denote lattice positions of the trimeric $a$-sites and monomeric
$b$-sites, respectively, which are occupied within the proposed Ising-Heisenberg model by 
the Heisenberg and Ising spins. Solid and broken lines schematically reproduce the intra-trimer Heisenberg interaction $J_{aa} = J_{\rm H} (\Delta)$ and the monomer-trimer Ising interaction 
$J_{ab} = J_{\rm I}$. The ellipse demarcates a six-spin cluster described by the Hamiltonian (\ref{eq2}), which can be mapped by the use of generalized star-triangle transformation 
into a simple triangle of the Ising spins mutually interacting via the effective exchange 
coupling $\beta J_{\rm kag}^{\rm eff}$.}
\label{fig1}
\end{figure*}

The aforementioned scientific achievements have stimulated an intensive search for inorganic molecular materials, whose paramagnetic metal centres connected in the crystal lattice via superexchange pathways 
would be strongly frustrated by their geometric arrangement.\cite{Gre01} From this perspective, 
the series of isostructural polymeric coordination compounds Cu$_9$X$_2$(cpa)$_6$.nH$_2$O (X=F, Cl, Br 
and cpa=carboxypentonic acid)\cite{Nor87} belongs to the most fascinating geometrically frustrated 
materials on behalf of a beautiful architecture of their magnetic lattice. The magnetic lattice of 
this series is built up of divalent copper (Cu$^{2+}$) ions situated at two crystalographically 
inequivalent lattice positions (see Fig.~\ref{fig1}). Cu$^{2+}$ ions with a square pyramidal
coordination ($a$-sites) form equilateral triangles (trimers), which are inter-connected to one 
another by Cu$^{2+}$ ions (monomers) with an elongated octahedral coordination ($b$-sites). 
The lattice positions of the $b$-sites constitute the regular Kagom\'e network, whereas 
each monomeric $b$-site is connected via four bonds to two adjacent trimers of the $a$-sites. 
This magnetic architecture can be accordingly regarded as the triangulated Kagom\'e 
(triangles-in-triangles) lattice, since smaller triangles of the trimeric $a$-sites are 
in fact embedded in larger triangles of the monomeric $b$-sites forming the basic Kagom\'e pattern.  

Experimental studies reported on this family of compounds reveal obvious manifestations of the geometric frustration. All three isomorphous compounds do not order down to 1.7~K,\cite{Mar94} 
the magnetization shows a plateau around one third of the saturation magnetization and it does not saturate up to 38~T.\cite{Mek98} Furthermore, the temperature dependence of inverse susceptibility indicates two temperature regimes inherent to two different exchange pathways. The linear dependence 
of the inverse susceptibility is well fitted by the Curie-Weiss law within the temperature interval between 150--250~K with the respective Weiss constants $\Theta_{\rm w}$ = -237~K, -226~K, and -243~K for the fluoro, chloro, and bromo analogue\cite{Mar94}, respectively, while the Weiss constant of the best Curie-Weiss fit generally increases up to roughly $\Theta_{\rm w}$ = 6~K in the temperature range below 50~K.\cite{Mek98} These observations would suggest an extremely high frustration ratio $f = |\Theta_{\rm w}|/T_{\rm c} > 130$ ($T_{\rm c}$ is the ordering temperature) \cite{Ram94} for each member of the Cu$_9$X$_2$(cpa)$_6$ family, which makes from this series a prominent class of the 
highly frustrated materials that possibly display the spin liquid ground state. Based on the considerations of exchange pathways, the strong antiferromagnetic interaction has been assigned to the exchange interaction $J_{aa}$ between the trimeric $a$-sites, while the weaker (possibly ferromagnetic) interaction has been ascribed to the exchange interaction $J_{ab}$ between the trimeric $a$-sites and the monomeric $b$-sites. The overall ratio between both exchange constants might be estimated from the corresponding ratio between the Weiss constants extrapolated from the high- and low-temperature regime yielding $|J_{aa}/J_{ab}| \approx 40$. 

Motivated by these experiments, Zheng and Sun\cite{Zhe05} have calculated an exact phase diagram 
of the spin-1/2 Ising model on the triangulated Kagom\'e lattice and the validity of this phase 
diagram has recently been confirmed by an independent exact calculation of Loh, Yao, and Carlson.\cite{Loh07} In addition to the exact ground-state and finite-temperature phase diagrams, 
the authors of the latter work also presented exact analytical results for several thermodynamic quantities (partition function, Helmholtz free energy, internal energy, entropy, 
and specific heat), which were complemented by the Monte Carlo simulations corroborating these 
exact analytical results and bringing other accurate numerical results for the magnetization and susceptibility in the zero as well as non-zero magnetic field.\cite{Loh07} Among the most 
interesting findings being reported is certainly a theoretical prediction of the spin liquid 
phase with a large residual entropy per spin $S_0/N_{\rm T} k_{\rm B} = \frac{1}{9} \ln 72 = 0.4752$, which appears in the ground state on assumption of a sufficiently strong antiferromagnetic 
intra-trimer interaction $J_{aa}/|J_{ab}| < -1$ (see for details Section IV.A in Ref.~\onlinecite{Loh07}). 

Unfortunately, the theoretical description based on the Ising model might fail in describing many 
important vestiges of the most of copper-based coordination compounds, since they usually exhibit 
a rather isotropic magnetic behavior.\cite{Jon01} Beside this, the Ising model description completely disregards the effect of quantum fluctuations, which might play a crucial role in determining 
the magnetic behavior of coordination compounds incorporating paramagnetic Cu$^{2+}$ ions having 
the lowest possible quantum spin number $S=1/2$. Therefore, the spin-1/2 quantum Heisenberg model 
is usually thought of as being much more appropriate model for the copper-based coordination compounds.\cite{Jon01} In accordance with this statement, ESR measurements performed on the Cu$_9$X$_2$(cpa)$_6$ series have shown just a minor anisotropy in the $g$-factor that also 
serves in evidence of a negligible magnetic anisotropy.\cite{Mek98,Oku98} 

Of course, the spin-1/2 quantum Heisenberg model is very difficult to deal with due to insurmountable 
mathematical complexities associated with a non-commutability between spin operators involved in its Hamiltonian and thus, one has usually to rely either on applicability of some simpler approximative method or to perform rather extensive numerical calculations. To the best of our knowledge, the 
spin-1/2 Heisenberg model on the triangulated Kagom\'e lattice has been studied yet merely in terms 
of the linear Holstein-Primakoff spin wave theory\cite{Nat97} and the variational mean-field like treatment.\cite{Str07} Both the aforementioned methods have however obvious insufficiencies. The former 
method based on the spin-wave approximation is applicable only if the ratio between both interaction 
parameters is from the interval $1 \leq |J_{aa}/J_{ab}| \leq 3$, i.e. the frustrating antiferromagnetic intra-trimer interaction $J_{aa}$ might not be much too stronger than the ferromagnetic one $J_{ab}$. 
On the other hand, the latter method based on the variational mean-field like treatment is asymptotically exact in the $|J_{ab}/J_{aa}| \to 0$ limit, but it even fails to predict the disordered ground state for any finite ratio $|J_{ab}/J_{aa}|$.\cite{Loh07}

In view of this, the present work aims to suggest and exactly solve the spin-1/2 Ising-Heisenberg 
model on the triangulated Kagom\'e lattice by establishing a precise mapping relationship with 
the corresponding spin-1/2 Ising model on the simple Kagom\'e lattice. Within the framework of the proposed Ising-Heisenberg model, the exchange interaction between the trimeric $a$-sites is being treated as the XXZ Heisenberg interaction $J_{aa} = J_{\rm H} (\Delta)$, while the exchange 
interaction between the trimeric $a$-sites and monomeric $b$-sites will be approximated by the Ising-type interaction $J_{ab} = J_{\rm I}$. Even although this model also has a clear deficiency 
in that the monomer-trimer interaction is being considered as the Ising-type interaction, 
it should be much more reliable in describing a frustrated magnetism of Cu$_9$X$_2$(cpa)$_6$ 
compounds as it correctly takes into account quantum fluctuations between the trimeric $a$-sites. 
In addition, it is also quite plausible to argue that the monomeric $b$-sites are in the frustrated regime completely free to flip without any energy cost and hence, the Ising character 
of the monomer-trimer interaction might be regarded as at least quite reasonable 
first-order approximation.  

The rest of this paper is organized as follows. In Section \ref{sec:model}, we will provide 
a detailed description of the model under investigation and then, basic steps of our exact 
analytical treatment will be explained. The most interesting results are presented and 
detailed discussed in Section \ref{sec:result}, where a particular emphasis is laid 
on a physical understanding of the ground-state and finite-temperature phase diagrams, 
as well as, the temperature dependences of several thermodynamical quantities. 
Some conclusions are finally drawn in Section \ref{sec:conc}. 

\section{\label{sec:model}Model and its exact solution}

Let us define the spin-1/2 Ising-Heisenberg model on the triangulated Kagom\'e lattice (Fig.~\ref{fig1}), which resembles a rather curious magnetic structure discovered in the series of polymeric coordination compounds Cu$_9$X$_2$(cpa)$_6$.\cite{Nor87} The magnetic properties of this series are captured to the lattice positions of Cu$^{2+}$ ions, which are situated at two 
inequivalent lattice positions previously referred to as the trimeric $a$-sites (empty circles) 
and the monomeric $b$-sites (full circles), respectively. Let us now assign the \textit{Heisenberg spin} $S=1/2$ to each trimeric $a$-site and the \textit{Ising spin} $\mu =1/2$ to each monomeric $b$-site.\footnote{To simplify further description, we will further refer to Cu$^{2+}$ ions 
at the trimeric and monomeric sites as to the Heisenberg and Ising spins, respectively.} 
In this way, the exchange interaction $J_{aa}$ between Cu$^{2+}$ ions located at the nearest-neighbor trimeric sites will be treated as the Heisenberg interaction $J_{\rm H}(\Delta)$, while the 
exchange interaction $J_{ab}$ between two Cu$^{2+}$ ions located at the nearest-neighbor 
trimeric and monomeric sites will be treated as the Ising interaction $J_{\rm I}$. 
The total Hamiltonian of the model under investigation then reads
\begin{eqnarray}
\hat{{\mathcal H}} = - J_{\rm H} \sum_{(i,j)} \Bigl[\Delta (\hat{S}_i^x \hat{S}_j^x + \hat{S}_i^y \hat{S}_j^y) + \hat{S}_i^z \hat{S}_j^z \Bigr]- J_{\rm I} \sum_{(k,l)} \hat{S}_k^z \hat{\mu}_l^z,     
\label{eq1}	
\end{eqnarray}
where the first summation is carried out over all pairs of the nearest-neighbor Heisenberg spins 
and the second summation extends over all pairs of the nearest-neighbor Heisenberg and Ising spins, respectively. The spin operators $\hat{S}_i^{\alpha} (\alpha = x, y, z)$ and $\hat{\mu}_l^z$ 
denote spatial components the usual spin-1/2 operator and the parameter $\Delta$ will allow 
us to control the exchange anisotropy in the anisotropic XXZ Heisenberg interaction and to obtain 
the Ising model as a special limiting case for $\Delta = 0$. Finally, the total number of 
the Ising spins ($b$-sites) is set to $N$, which implies that the total number of all spins 
(lattice sites) is $N_{\rm T} = 3N$.

For further convenience, the total Hamiltonian (\ref{eq1}) can be written as a sum over six-spin
cluster Hamiltonians $\hat{{\mathcal H}} = \sum_{k} \hat{{\mathcal H}}_k$, whereas each cluster 
Hamiltonian $\hat{{\mathcal H}}_k$ involves all the interaction terms associated with 
three Heisenberg spins from $k$th trimer (see Fig.~\ref{fig1})
\begin{widetext}
\begin{eqnarray}
\hat{{\mathcal H}}_k = &-& J_{\rm H} \left[\Delta (\hat{S}_{k1}^x \hat{S}_{k2}^x 
                        + \hat{S}_{k1}^y \hat{S}_{k2}^y) + \hat{S}_{k1}^z \hat{S}_{k2}^z \right] 
                        - J_{\rm H} \left[\Delta (\hat{S}_{k2}^x \hat{S}_{k3}^x 
                        + \hat{S}_{k2}^y \hat{S}_{k3}^y) + \hat{S}_{k2}^z \hat{S}_{k3}^z \right] \nonumber \\
                       &-& J_{\rm H} \left[\Delta (\hat{S}_{k3}^x \hat{S}_{k1}^x 
                        + \hat{S}_{k3}^y \hat{S}_{k1}^y) + \hat{S}_{k3}^z \hat{S}_{k1}^z \right] 
                        - \hat{S}_{k1}^z (h_{k1} + h_{k2}) - \hat{S}_{k2}^z (h_{k2} + h_{k3}) 
                        -  \hat{S}_{k3}^z (h_{k3} + h_{k1}),
\label{eq2}	                        
\end{eqnarray}
\end{widetext}
where $h_{ki} = J_{\rm I} \hat{\mu}_{ki}^z$. Obviously, the six-spin cluster Hamiltonian (\ref{eq2})
might be regarded as the Hamiltonian of $k$th Heisenberg trimer placed in some generally non-uniform 
magnetic field produced by three surrounding Ising spins. By taking into account the commutability between different cluster Hamiltonians $[\hat{{\mathcal H}}_i, \hat{{\mathcal H}}_j] = 0$ valid for each $i \neq j$, the partition function of the spin-1/2 Ising-Heisenberg model on the triangulated Kagom\'e lattice can be partially factorized into the product of cluster partition functions 
${\mathcal Z}_k$
\begin{eqnarray}
{\mathcal Z} = \sum_{\{ \mu_i \}} \prod_{k=1}^{2N/3} \mbox{Tr}_{k} \exp(- \beta \hat{{\mathcal H}}_k) 
             = \sum_{\{ \mu_i \}} \prod_{k=1}^{2N/3} {\mathcal Z}_k. 
\label{eq3}	
\end{eqnarray}
Above, the summation $\sum_{\{ \mu_i \}}$ runs over all possible spin configurations of the Ising spins
and the symbol $\mbox{Tr}_{k}$ stands for a trace over spin degrees of freedom of $k$th Heisenberg trimer. To proceed further with the calculation, it is very convenient to accomplish an exact analytical diagonalization of the cluster Hamiltonian (\ref{eq2}) in a particular Hilbert subspace corresponding to $k$th Heisenberg trimer. If doing so, the resulting cluster partition function ${\mathcal Z}_k$ will depend just upon three Ising spins $\mu_{k1}$, $\mu_{k2}$, and $\mu_{k3}$, 
which are included in the parameters $h_{k1}$, $h_{k2}$, and $h_{k3}$ determining the 'local magnetic fields' acting on the Heisenberg spins. Moreover, the explicit mathematical form of the cluster partition function ${\mathcal Z}_k$ immediately implies a possibility of performing the generalized star-triangle mapping transformation\cite{Fis59,Syo72} 
\begin{widetext}
\begin{eqnarray}
{\mathcal Z}_k &=& 2 \exp \left(\frac{3 \beta J_{\rm H}}{4} \right) 
    \cosh \left[\beta J_{\rm I}(\mu_{k1}^z + \mu_{k2}^z + \mu_{k3}^z) \right] \nonumber \\
  &+& \exp \left[- \frac{\beta J_{\rm H}}{4} + \frac{\beta J_{\rm I}}{3}  (\mu_{k1}^z 
   + \mu_{k2}^z + \mu_{k3}^z) \right] \sum_{n=0}^2 \exp \left[-2 \beta \mbox{sgn}(Q_{+}) \sqrt{P} 
   \cos \left(\Phi_{+} + \frac{2 \pi n}{3} \right) \right]  \nonumber \\
&+& \exp \left[- \frac{\beta J_{\rm H}}{4} - \frac{\beta J_{\rm I}}{3} (\mu_{k1}^z + \mu_{k2}^z 
+ \mu_{k3}^z) \right] \sum_{n=0}^2 \exp \left[-2 \beta \mbox{sgn}(Q_{-}) \sqrt{P} 
\cos \left(\Phi_{-} + \frac{2 \pi n}{3} \right) \right] \nonumber \\
&=& A \exp \left[ \beta J_{\rm kag}^{\rm eff} \left(\mu_{k1}^z \mu_{k2}^z + \mu_{k2}^z \mu_{k3}^z
 + \mu_{k3}^z \mu_{k1}^z \right) \right],
\label{eq4}
\end{eqnarray}
\end{widetext}
where $\beta = 1/(k_{\rm B} T)$, $k_{\rm B}$ is Boltzmann's constant, $T$ the absolute temperature,
and the parameters $P$, $Q_{\pm}$, and $\Phi_{\pm}$ are defined as follows
\begin{eqnarray}
P &=& \left(\frac{J_{\rm I}}{3} \right)^2 \left[\frac{3}{4} - (\mu_{k1}^z \mu_{k2}^z 
+ \mu_{k2}^z \mu_{k3}^z  + \mu_{k3}^z \mu_{k1}^z) \right] \nonumber \\
&+& \left(\frac{J_{\rm H} \Delta}{2} \right)^2,  \label{eq5a}	\\
Q_{\pm} &=& \pm \frac{1}{2} \left(\frac{J_{\rm I}}{3} \right)^3 \left[\mu_{k1}^z  
         + \mu_{k2}^z + \mu_{k3}^z - 12 \mu_{k1}^z \mu_{k2}^z \mu_{k3}^z \right] \nonumber \\
        &-& \left(\frac{J_{\rm H} \Delta}{2} \right)^3, \label{eq5b}	\\
\Phi_{\pm} &=& \frac{1}{3} \arctan \left(\frac{\sqrt{P^3 - Q_{\pm}^2}}{Q_{\pm}} \right). \label{eq5c}	
\end{eqnarray}
From here onward, our procedure closely follows the approach developed by Zheng and Sun that 
relates an exact solution of the spin-1/2 Ising model on the triangulated Kagom\'e lattice 
to an exact solution of the the corresponding spin-1/2 Ising model on the simple Kagom\'e lattice.\cite{Zhe05} As a matter of fact, the star-triangle transformation (\ref{eq4}) effectively removes all the interaction terms associated with $k$th Heisenberg trimer and substitutes them 
by the effective interaction $J_{\rm kag}^{\rm eff}$ between three enclosing Ising spins $\mu_{k1}$, $\mu_{k2}$, and $\mu_{k3}$. Of course, the mapping transformation (\ref{eq4}) must hold for any 
available spin configuration of three enclosing Ising spins and this self-consistency condition unambiguously determines so far not specified mapping parameters $A$ and $\beta J_{\rm kag}^{\rm eff}$
\begin{eqnarray}
A &=& \left(V_1 V_2^3 \right)^{1/4} \quad \mbox{and} \quad 
\beta J_{\rm kag}^{\rm eff} = \ln \left(V_1/V_2 \right). 
\label{eq6}	
\end{eqnarray}
The functions $V_1$ and $V_2$ entering into the effective mapping parameters $A$ and 
$\beta J_{\rm kag}^{\rm eff}$ are actually two independent expressions for the cluster 
partition function (\ref{eq4}) to be obtained by considering all eight possible 
spin configurations available to three enclosing Ising spins 
\begin{widetext}
\begin{eqnarray}
V_1 &=& 2 \exp \left(\frac{3 \beta J_{\rm H}}{4} \right) \cosh \left(\frac{3\beta J_{\rm I}}{2} \right) 
     +  2 \exp \left(- \frac{\beta J_{\rm H}}{4} \right) \cosh \left(\frac{\beta J_{\rm I}}{2} \right)
        \left[\exp \left(\beta J_{\rm H} \Delta \right) 
     + 2 \exp \left(-\frac{\beta J_{\rm H} \Delta}{2} \right)\right], \label{eq7a} \\
V_2 &=& 2 \exp \left(\frac{3 \beta J_{\rm H}}{4} \right) \cosh \left(\frac{\beta J_{\rm I}}{2} \right) 
     + \exp \left(- \frac{\beta J_{\rm H}}{4} + \frac{\beta J_{\rm I}}{6} \right) \sum_{n=0}^2 
     \exp \left[-2 \beta \mbox{sgn}(q_{+}) \sqrt{p} \cos \left(\phi_{+} + \frac{2 \pi n}{3} \right) \right]  \nonumber \\
&+& \exp \left(- \frac{\beta J_{\rm H}}{4} - \frac{\beta J_{\rm I}}{6} \right) \sum_{n=0}^2 
    \exp \left[-2 \beta \mbox{sgn}(q_{-}) \sqrt{p} \cos \left(\phi_{-} + \frac{2 \pi n}{3} \right) \right], 
\label{eq7b}	
\end{eqnarray}
\end{widetext}
and the parameters $p$, $q_{\pm}$, and $\phi_{\pm}$ are defined as follows
\begin{eqnarray}
p &=& \left(\frac{J_{\rm I}}{3} \right)^2 + \left(\frac{J_{\rm H} \Delta}{2} \right)^2, \label{eq8a} \\
q_{\pm} &=& \pm \left(\frac{J_{\rm I}}{3} \right)^3 - \left(\frac{J_{\rm H} \Delta}{2} \right)^3, \label{eq8b} \\
\phi_{\pm} &=& \frac{1}{3} \arctan \left(\frac{\sqrt{p^3 - q_{\pm}^2}}{q_{\pm}} \right). \label{eq8c}	
\end{eqnarray} 
By substituting the mapping transformation (\ref{eq4}) into the relevant expression for the partition 
function (\ref{eq3}) one establishes a simple mapping relationship between the partition function 
of the spin-1/2 Ising-Heisenberg model on the triangulated Kagom\'e lattice and the partition function 
of the corresponding spin-1/2 Ising model on the Kagom\'e lattice 
\begin{eqnarray}
{\mathcal Z} (\beta, J_{\rm H}, \Delta, J_{\rm I}) 
  = A^{2N/3} {\mathcal Z}_{\rm kag} (\beta J_{\rm kag}^{\rm eff}). 
\label{eq9}	
\end{eqnarray}
This relation actually completes our exact calculation of the partition function as the corresponding 
exact result for the partition function of the spin-1/2 Ising model on the Kagom\'e lattice is well 
known.\cite{Syo51,Kan53,Dom60} At this stage, exact results for other thermodynamic quantities follow straightforwardly. The Helmholtz free energy of the spin-1/2 Ising-Heisenberg model on the 
triangulated Kagom\'e lattice may be connected to the Helmholtz free energy of the corresponding 
spin-1/2 Ising model on the Kagom\'e lattice 
(${\mathcal F}_{\rm kag} = - \beta^{-1} \ln {\mathcal Z}_{\rm kag}$) through the relation
\begin{eqnarray}
{\mathcal F} = {\mathcal F}_{\rm kag} - 2N \beta^{-1} \ln A /3.  
\label{eq10}	
\end{eqnarray}
The connection between the internal energy of spin-1/2 Ising-Heisenberg model on the triangulated 
Kagom\'e lattice and that one of the corresponding spin-1/2 Ising model on the Kagom\'e lattice 
can readily be obtained by differentiating the logarithm of Eq.~(\ref{eq9}) with respect to 
the inverse temperature $\beta$ yielding
\begin{eqnarray}
{\mathcal U} &=& - \frac{\partial \ln {\mathcal Z}}{\partial \beta} 
              = - \frac{2N}{3} \frac{\partial \ln A}{\partial \beta} 
                - \frac{\partial \ln {\mathcal Z}_{\rm kag}}{\partial (\beta J^{\rm eff}_{\rm kag})} 
                  \frac{\partial (\beta J_{\rm kag}^{\rm eff})}{\partial \beta} \nonumber \\
 &=& \frac{W_1}{V_1} \left(\frac{{\mathcal U}_{\rm kag}}{J^{\rm eff}_{\rm kag}} - \frac{N}{6} \right) 
  - \frac{W_2}{V_2} \left(\frac{{\mathcal U}_{\rm kag}}{J^{\rm eff}_{\rm kag}} + \frac{N}{2} \right).  
\label{eq11}	
\end{eqnarray}
Note that the exact result for the internal energy of the spin-1/2 Ising model on the Kagom\'e lattice
is well known (see for instance Ref.~\onlinecite{Dom60}) and thus, our exact calculation is essentially completed by introducing the parameters $W_1$ and $W_2$ that denote inverse temperature derivatives 
of the functions $V_1$ and $V_2$ given by Eqs.~(\ref{eq7a}) and (\ref{eq7b}) 
\begin{widetext}
\begin{eqnarray}
W_1 &=& \frac{\partial V_1}{\partial \beta} 
     = \frac{3}{2} \exp \left(\frac{3 \beta J_{\rm H}}{4} \right) 
          \left[J_{\rm H} \cosh \left(\frac{3 \beta J_{\rm I}}{2} \right) 
        + 2 J_{\rm I} \sinh \left(\frac{3 \beta J_{\rm I}}{2} \right) \right] \nonumber \\
    &+& \exp \left(-\frac{\beta J_{\rm H}}{4} + \beta J_{\rm H} \Delta \right) 
    \left[\left(2 J_{\rm H} \Delta - \frac{J_{\rm H}}{2} \right) \cosh \left(\frac{\beta J_{\rm I}}{2}  
    \right) + J_{\rm I} \sinh \left(\frac{\beta J_{\rm I}}{2} \right) \right] \nonumber \\
    &-& \exp \left(-\frac{\beta J_{\rm H}}{4} - \frac{\beta J_{\rm H} \Delta}{2} \right) 
    \left[\left(2 J_{\rm H} \Delta + J_{\rm H} \right) \cosh \left(\frac{\beta J_{\rm I}}{2} \right) 
     - 2 J_{\rm I} \sinh \left(\frac{\beta J_{\rm I}}{2} \right) \right],  \label{eq12a} \\
W_2 &=& \frac{\partial V_2}{\partial \beta} 
     =  \exp \left(\frac{3 \beta J_{\rm H}}{4} \right) \left[\frac{3}{2} J_{\rm H} 
       \cosh \left(\frac{\beta J_{\rm I}}{2} \right) 
         + J_{\rm I} \sinh \left(\frac{\beta J_{\rm I}}{2} \right) \right]  \\ 
        &-&  \sum_{n=0}^2 \left[ \frac{J_{\rm H}}{4} - \frac{J_{\rm I}}{6} 
        + 2 \mbox{sgn}(q_{+}) \sqrt{p} \cos \left(\phi_{+} + \frac{2 \pi n}{3} \right) \right] 
      \exp \left[- \frac{\beta J_{\rm H}}{4} + \frac{\beta J_{\rm I}}{6} - 2 \beta \mbox{sgn}(q_{+}) 
          \sqrt{p} \cos \left(\phi_{+} + \frac{2 \pi n}{3} \right) \right]  \nonumber \\ 
       &-&  \sum_{n=0}^2 \left [ \frac{J_{\rm H}}{4} + \frac{J_{\rm I}}{6} + 
        2 \mbox{sgn}(q_{-}) \sqrt{p} \cos \left(\phi_{-} + \frac{2 \pi n}{3} \right) \right ] 
      \exp \left[- \frac{\beta J_{\rm H}}{4} - \frac{\beta J_{\rm I}}{6} - 2 \beta \mbox{sgn}(q_{-}) 
          \sqrt{p} \cos \left(\phi_{-} + \frac{2 \pi n}{3} \right) \right]. \nonumber
\label{eq12b}	
\end{eqnarray}
\end{widetext}
The entropy can be now easily calculated either from the basic thermodynamic relation between Helmholtz free energy and internal energy ${\mathcal F} = {\mathcal U} - T {\mathcal S}$, or as a negative temperature derivative of the free energy (\ref{eq10}). Both procedures yield the following closed-form relation for the reduced entropy per one site of the original spin-1/2 Ising-Heisenberg model on the triangulated Kagom\'e lattice 
\begin{eqnarray}
\frac{{\mathcal S}}{N_{\rm T} k_{\rm B}} = \frac{1}{3N} \ln {\mathcal Z}_{\rm kag} &+& \frac{2}{9} \ln A + \frac{\beta W_1}{3 V_1} \left(\frac{{\mathcal U}_{\rm kag}}{N J^{\rm eff}_{\rm kag}} 
- \frac{1}{6} \right) \nonumber \\
&-& \frac{\beta W_2}{3 V_2} 
\left(\frac{{\mathcal U}_{\rm kag}}{N J^{\rm eff}_{\rm kag}} + \frac{1}{2} \right).  
\label{eq13}	
\end{eqnarray}
It is quite obvious from the above formula that the reduced entropy of the spin-1/2 Ising-Heisenberg model on the triangulated Kagom\'e lattice can be expressed in terms of the known exact results 
for the partition function \cite{Kan53} and internal energy \cite{Dom60} of the corresponding 
spin-1/2 Ising model on the Kagom\'e lattice. It is worthy to note, moreover, that the zero-field specific heat can also be simply obtained as a temperature derivative of the internal energy (\ref{eq11}), but the final expression is too cumbersome to write it down here explicitly.

\section{\label{sec:result}Results and discussion}

Before proceeding to a discussion of the most interesting results, it is quite useful to realize 
that all final results derived in the foregoing section are invariant under the transformation 
$J_{\rm I} \to -J_{\rm I}$. For this reason, it is very convenient to set the absolute value 
of the Ising interaction $|J_{\rm I}|$ as the energy unit and to define two dimensionless parameters 
$k_{\rm B} T/|J_{\rm I}|$ and $J_{\rm H}/|J_{\rm I}|$ reducing the number of free parameters. 
The former dimensionless parameter is then proportional to a relative size of the temperature, 
while the latter one determines a relative strength of the intra-trimer Heisenberg interaction 
$J_{\rm H}$ with respect to the monomer-trimer Ising interaction $J_{\rm I}$. 

\begin{figure}
\begin{center}
\includegraphics[width=7cm]{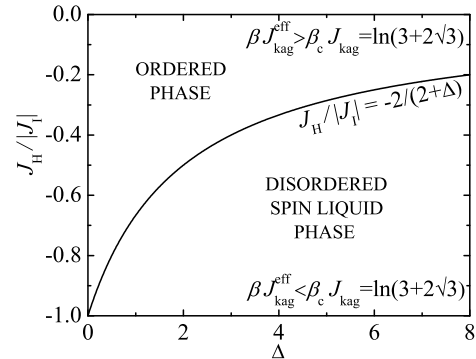} 
\end{center} 
\vspace{-11mm} 
\caption{Ground-state phase diagram in the $\Delta-J_{\rm H}/|J_{\rm I}|$ plane 
separating the spontaneously ordered and disordered phases.}
\label{fig2}
\end{figure}

First, let us take a closer look at the ground-state behavior. It is quite evident that the 
spin-1/2 Ising-Heisenberg model on the triangulated Kagom\'e lattice exhibits spontaneous
long-range ordering if and only if the corresponding spin-1/2 Ising model on the simple 
Kagom\'e lattice is spontaneously long-range ordered as well. Accordingly, the ground-state 
phase diagram can readily be obtained from a comparison of the effective interaction parameter 
$\beta J_{\rm kag}^{\rm eff}$ given by Eq.~(\ref{eq6}) with the critical point of the spin-1/2 
Ising model on the Kagom\'e lattice $\beta_{\rm c} J_{\rm kag} = \ln(3 + 2 \sqrt{3})$ 
[$\beta_{\rm c} = 1/(k_{\rm B} T_{\rm c})$, $T_{\rm c}$ is the critical temperature].\cite{Syo51} 
In the consequence of that, the ground state is spontaneously long-range ordered if 
$\beta J_{\rm kag}^{\rm eff} > \ln(3 + 2 \sqrt{3})$, while it becomes disordered spin liquid 
state as long as $\beta J_{\rm kag}^{\rm eff} < \ln (3 + 2 \sqrt{3})$. The phase boundary 
between ordered and disordered ground states, which follows from the zero-temperature limit 
of the mapping parameter (\ref{eq6}), is shown in Fig.~\ref{fig2} and can be expressed through 
the following analytical condition 
\begin{eqnarray}
J_{\rm H}/|J_{\rm I}| = -2/(2 + \Delta). 
\label{eq14}	
\end{eqnarray}
It can be easily understood that the disordered spin liquid state appears as a result of geometric frustration, which comes into play provided that there is a sufficiently strong antiferromagnetic Heisenberg interaction. Indeed, the ground state is a simple ferromagnetic or 
ferrimagnetic spin arrangement for any $J_{\rm H}/|J_{\rm I}|$ greater than the boundary value (\ref{eq14}) depending on whether $J_{\rm I} > 0$ or $J_{\rm I} < 0$, otherwise it becomes 
the disordered spin liquid state. Apparently, the greater is the anisotropy constant $\Delta$, 
the weaker antiferromagnetic Heisenberg interaction  is needed to destroy 
the spontaneous long-range ordering. 

Now, let us make few remarks on a nature of possible ground states. The ferromagnetic as well as ferrimagnetic ordered states can be characterized through the same classical spin ordering as previously described by the analysis of the spin-1/2 Ising model on the triangulated Kagom\'e lattice.\cite{Loh07} As far as the disordered spin liquid state is concerned, however, there is 
a fundamental difference between the disordered ground state of the spin-1/2 Ising-Heisenberg 
model with any $\Delta \neq 0$ and its semi-classical Ising limit $\Delta = 0$, respectively. 
As a matter of fact, considering the frustrated regime $J_{\rm H}/|J_{\rm I}| < -1$ and setting 
$\Delta = 0$ into Eq.~(\ref{eq6}) gives the zero effective exchange coupling 
$\beta J_{\rm kag}^{\rm eff} = 0$, which means that the disordered ground state of the spin-1/2 
Ising model on the triangulated Kagom\'e lattice is equivalent to an ensemble of non-interacting 
spins, or equivalently, to a spin system at infinite temperature. This implies that the Ising spins 
at the monomeric $b$-sites are completely free to flip and the ground-state degeneracy of the disordered spin liquid state can be straightforwardly counted by following the argumentation of 
Loh, Yao, and Carlson.\cite{Loh07} The residual entropy per spin is accordingly 
$S_0/N_{\rm T} k_{\rm B} = \frac{1}{9} \ln 72 = 0.4752$, since each trimeric unit has precisely 
three different lowest-energy states for each possible spin configuration of its three surrounding monomeric spins and the basic unit cell contains two trimeric units and three monomeric spins. 

Contrary to this, the effective exchange coupling (\ref{eq6}) that corresponds to the spin-1/2 Ising-Heisenberg model on the triangulated Kagom\'e lattice with any $\Delta \neq 0$ is equal 
to $\beta J_{\rm kag}^{\rm eff} = \ln 2$ in the disordered region. Even although this value 
is smaller than the critical value $\beta_{\rm c} J_{\rm kag}$ and the spin system must be 
consequently disordered, its positive and non-zero value indicates a ferromagnetic character 
of short-range correlations between the monomeric Ising spins. It actually turns out that the lowest-energy state of each Heisenberg trimer is two-fold degenerate if all three surrounding 
Ising spins are aligned alike (all three Ising spins point either 'up' or 'down'), while there 
is just one lowest-energy state from the ground-state manifold provided that two from three 
Ising spins are aligned alike and the third spin points in the opposite direction (remember 
that the lowest-energy state of the trimeric unit is three-fold degenerate irrespective of 
a spin configuration of monomeric spins within the Ising model). With regard to this, the residual 
entropy of the disordered spin liquid phase is for the quantum Ising-Heisenberg model significantly lower than for its semi-classical Ising limit $S_0/N_{\rm T} k_{\rm B} = 0.2806$ and $0.4752$, respectively, which implies that quantum fluctuations partially lift a macroscopic degeneracy 
of the ground-state manifold. The most surprising finding stemming from our study is that 
the zero-point entropy of the spin-1/2 Ising-Heisenberg model is the same for arbitrary but 
non-zero anisotropy $\Delta$ and this observation suggests that the semi-classical Ising limit 
represents a very special limiting case of the model under consideration.

\begin{figure}
\begin{center}
\includegraphics[width=7cm]{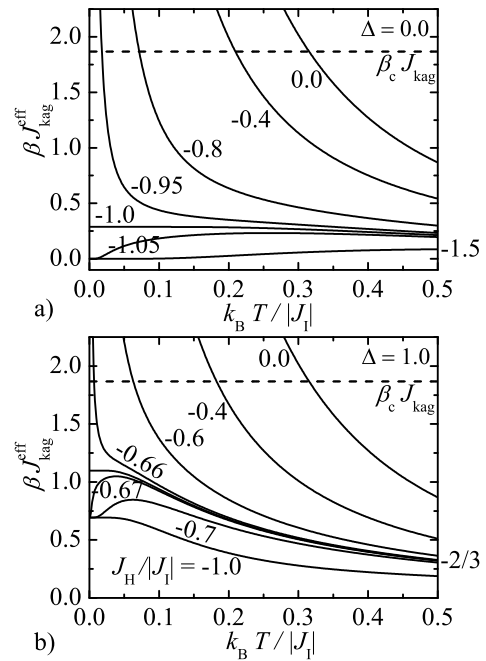} 
\end{center} 
\vspace{-11mm} 
\caption{Temperature dependences of the effective interaction $\beta J_{\rm kag}^{\rm eff}$ 
for several values of the ratio $J_{\rm H}/|J_{\rm I}|$ and two different anisotropy constants: 
(a) $\Delta = 0$; (b) $\Delta = 1$. Broken line shows the critical point $\beta_{\rm c} J_{\rm kag} = \ln (3 + 2 \sqrt{3})$ of the corresponding spin-1/2 Ising model on the Kagom\'e lattice.}
\label{fig3}
\end{figure}

To provide an independent check of the aforementioned scenario, we depict in Fig.~\ref{fig3} temperature variations of the effective interaction $\beta J_{\rm kag}^{\rm eff}$ for several 
values of the ratio $J_{\rm H}/|J_{\rm I}|$ and two different values of the anisotropy 
constant $\Delta$ = 0 and 1, respectively. 
It should be mentioned that the particular case with $\Delta = 1$ qualitatively resembles 
temperature dependences of the mapping parameter $\beta J_{\rm kag}^{\rm eff}$ for the 
Ising-Heisenberg model with any 
$\Delta \neq 0$. The considered model system is spontaneously long-range ordered 
below certain critical temperature if and only if the effective interaction parameter 
$\beta J_{\rm kag}^{\rm eff}$ is greater than the critical value $\beta_{\rm c} J_{\rm kag} 
= \ln (3 + 2 \sqrt{3})$ shown in Fig.~\ref{fig3} as a broken line. In agreement with 
the aforedescribed ground-state analysis, the effective interaction starts from zero, 
then gradually increases to some local maximum before it finally goes to zero by increasing 
temperature whenever $\Delta = 0$ and $J_{\rm H}/|J_{\rm I}| < -1$. On the other hand, 
the effective interaction generally starts from the value $\ln 2$, then exhibits 
a temperature-induced local maximum before it finally tends to zero whenever $\Delta \neq 0$ 
and $J_{\rm H}/|J_{\rm I}| < -2/(2 + \Delta)$. It is also quite evident from Fig.~\ref{fig3} 
that the spin system remains disordered over the whole temperature range for any 
$J_{\rm H}/|J_{\rm I}| < -2/(2 + \Delta)$, since the effective interaction 
$\beta J_{\rm kag}^{\rm eff}$ never crosses the critical value $\beta_{\rm c} J_{\rm kag}$ 
that is needed to invoke a spontaneous ordering. This observation would suggest that the Ising-Heisenberg model in question cannot exhibit a temperature-induced reentrant phase 
transition from the disordered towards the spontaneously ordered phase regardless of 
whether $\Delta = 0$ or $\Delta \neq 0$.

\begin{figure}
\begin{center}
\includegraphics[width=7cm]{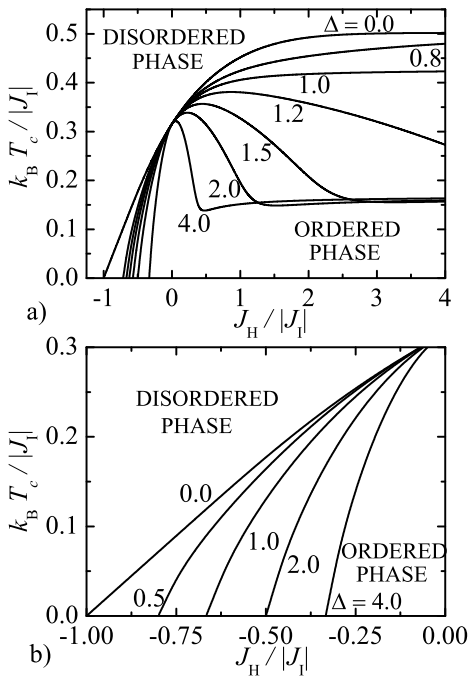} 
\end{center} 
\vspace{-11mm} 
\caption{The dimensionless critical temperature $k_{\rm B} T_{\rm c}/|J_{\rm I}|$ as a function 
of the ratio $J_{\rm H}/|J_{\rm I}|$ for different values of the exchange anisotropy $\Delta$.
Fig.~\ref{fig4}b shows a detail of the finite-temperature phase diagram, where the critical 
temperature tends towards the zero temperature.}
\label{fig4}
\end{figure}

At this point, let us proceed to a discussion of the finite-temperature phase diagram. The critical temperature of the spin-1/2 Ising-Heisenberg model on the triangulated Kagom\'e lattice can easily 
be calculated from the critical condition $\beta_{\rm c} J_{\rm kag}^{\rm eff} = \ln (3 + 2 \sqrt{3})$, which ensures that the corresponding spin-1/2 Ising model on the Kagom\'e lattice is precisely 
at the critical point. Note that this critical condition is essentially equivalent to a graphical solution that finds points of intersection between the temperature dependence of the effective interaction $\beta J_{\rm kag}^{\rm eff}$ and the critical point $\beta_{\rm c} J_{\rm kag} 
= \ln (3 + 2 \sqrt{3})$ of the spin-1/2 Ising model on the Kagom\'e lattice (see Fig.~\ref{fig3}). 
The dependence of dimensionless critical temperature $k_{\rm B} T_{\rm c}/|J_{\rm I}|$ on 
the ratio $J_{\rm H}/|J_{\rm I}|$ is displayed in Fig.~\ref{fig4} for several values of 
the anisotropy parameter $\Delta$.  
As one can see, the critical temperature monotonically decreases by decreasing the ratio 
$J_{\rm H}/|J_{\rm I}|$ until it vanishes at the ground-state phase boundary (\ref{eq14})
whenever the easy-axis exchange anisotropies $\Delta<1$ are assumed. By contrast, the interesting non-monotonic dependence of the critical temperature may be observed for the easy-plane 
anisotropies $\Delta>1$ presumably due to a competition between the easy-axis Ising interaction 
$J_{\rm I}$ and the easy-plane Heisenberg interaction $J_{\rm H} (\Delta)$ before the critical 
line finally tends to zero-temperature limit consistent with the ground-state boundary (\ref{eq14}). Several limiting cases of the model under investigation can be checked. First, the critical line 
of the spin-1/2 Ising model on the triangulated Kagom\'e lattice originally reported by 
Zheng and Sun\cite{Zhe05} and later rederived by Loh, Yao, and Carlson,\cite{Loh07} 
is recovered on assumption that $\Delta = 0$. By assuming $J_{\rm H}/|J_{\rm I}|=0$, 
on the other hand, all critical lines meet at a common critical point  
\begin{eqnarray}
\frac{k_{\rm B} T_{\rm c}}{|J_{\rm I}|} 
     = \frac{1}{2 \ln \left(\sqrt{3 + 2 \sqrt{3}} + \sqrt{2+ 2 \sqrt{3}}\right)} \doteq 0.3154,
\label{eq15}	
\end{eqnarray}
which is consistent with the critical temperature of the spin-1/2 Ising model 
on the decorated Kagom\'e lattice. It is also quite interesting to check the asymptotic behavior 
of the critical temperature achieved in the limit $J_{\rm H}/|J_{\rm I}| \to \infty$ that 
corresponds to the infinitely strong ferromagnetic Heisenberg interaction. If $\Delta < 1$, 
the critical temperature asymptotically reaches the value
\begin{eqnarray}
\frac{k_{\rm B} T_{\rm c}}{|J_{\rm I}|} 
     = \frac{1}{\ln \left(2 + \sqrt{3} + \sqrt{6 + 4 \sqrt{3}} \right)} \doteq 0.5021,
\label{eq16}	
\end{eqnarray} 
while for $\Delta > 1$ it strikingly decreases down to one third of this asymptotic value
\begin{eqnarray}
\frac{k_{\rm B} T_{\rm c}}{|J_{\rm I}|} 
     = \frac{1}{3 \ln \left(2 + \sqrt{3} + \sqrt{6 + 4 \sqrt{3}} \right)} \doteq 0.1673.
\label{eq17}	
\end{eqnarray} 
For the particular case $\Delta = 1$, the critical temperature of the spin-1/2 Ising-Heisenberg 
model on the triangulated Kagom\'e lattice acquires in the limit $J_{\rm H}/|J_{\rm I}| \to \infty$
the asymptotic value $k_{\rm B} T_{\rm c}/|J_{\rm I}| \doteq 0.4285$.
 
\begin{figure}
\begin{center}
\includegraphics[width=6.75cm]{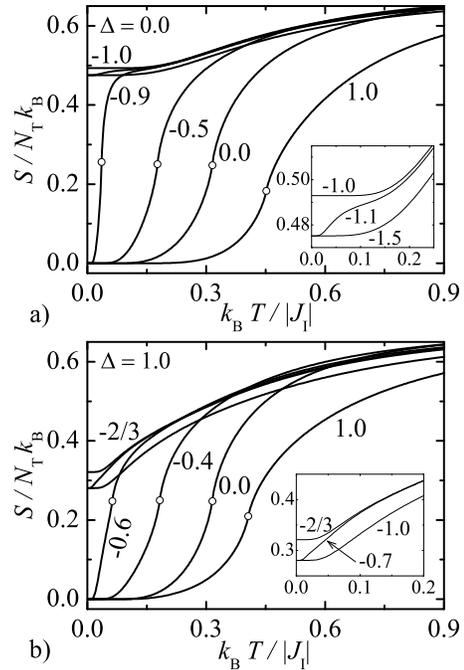} 
\end{center} 
\vspace{-11mm} 
\caption{Thermal variations of the entropy for different values of the ratio $J_{\rm H}/|J_{\rm I}|$,
which are indicated by the numbers associated with each line, and two different values of the anisotropy constant: (a) $\Delta = 0$; (b) $\Delta = 1$. The insert shows the low-temperature 
behavior for special cases, where the entropy tends towards its residual values.}
\label{fig5}
\end{figure}

Next, let us turn our attention to temperature dependences of some thermodynamic quantities.
Fig.~\ref{fig5} depicts temperature variations of the entropy calculated for several values 
of the ratio $J_{\rm H}/|J_{\rm I}|$ and two different values of the anisotropy $\Delta$ = 0 and 1, respectively. As it can be clearly seen, the standard S-shaped dependence with a weak energy-type singularity located at critical points of the order-disorder phase transitions, which are schematically 
shown in Fig.~\ref{fig5} as open circles, gradually shifts towards zero temperature upon strengthening the antiferromagnetic Heisenberg interaction until the weak singularity completely disappears 
from the entropy dependence whenever $J_{\rm H}/|J_{\rm I}|<-2/(2 + \Delta)$. Under this circumstance, the displayed thermal dependences tend asymptotically towards the zero-point entropy 
$S_0/N_{\rm T} k_{\rm B} = 0.2806$ or $0.4752$ depending on whether $\Delta \neq 0$ or $\Delta = 0$, respectively. The low-temperature behavior of the entropy, which is for better clarity shown 
in the insert of Fig.~\ref{fig5}, thus provides an independent confirmation of the macroscopic degeneracy that appears within the disordered spin liquid ground states.  

\begin{figure*}
\begin{center}
\includegraphics[width=15cm]{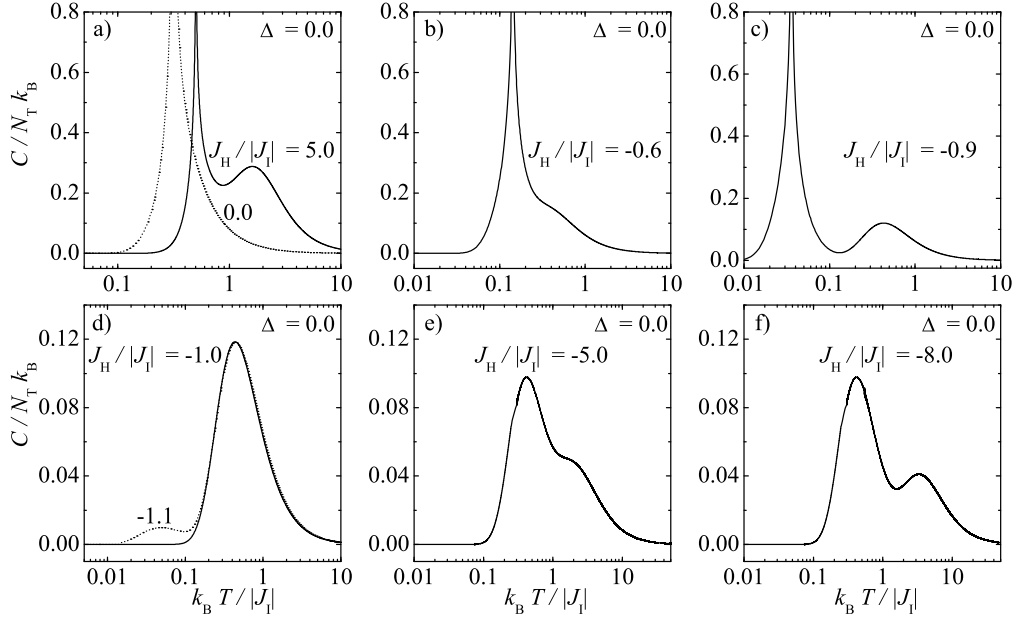} 
\end{center} 
\vspace{-16mm} 
\caption{Some typical temperature dependences of the zero-field specific heat of the spin-1/2 Ising model on the triangulated Kagom\'e lattice with $\Delta = 0$, which can be obtained by changing 
a strength of the interaction ratio $J_{\rm H}/|J_{\rm I}|$.}
\label{fig6}
\end{figure*}

\begin{figure*}
\begin{center}
\includegraphics[width=15cm]{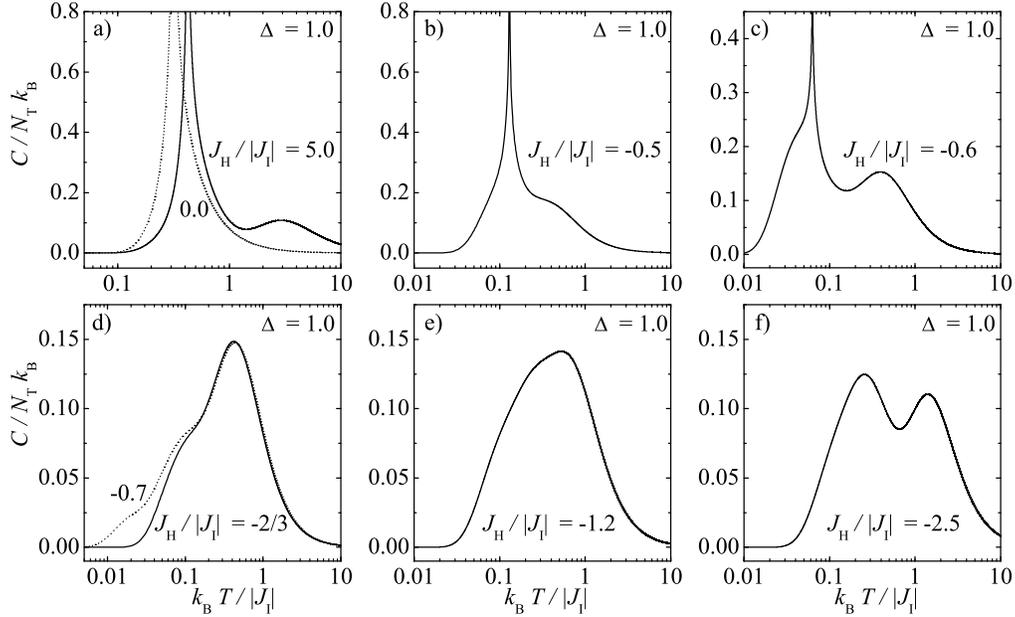} 
\end{center} 
\vspace{-16mm} 
\caption{Some typical temperature dependences of the zero-field specific heat of the spin-1/2 Ising-Heisenberg model on the triangulated Kagom\'e lattice with $\Delta = 1$, which can be 
obtained by changing a strength of the interaction ratio $J_{\rm H}/|J_{\rm I}|$.}
\label{fig7}
\end{figure*}

Finally, let us conclude our analysis of thermodynamics by exploring temperature dependences 
of the zero-field specific heat. For illustration, some typical thermal variations of the 
specific heat are plotted in Figs.~\ref{fig6} and \ref{fig7} for several values of the ratio 
$J_{\rm H}/|J_{\rm I}|$ and two different values of the exchange anisotropy $\Delta$ = 0 
and 1, respectively. The upper panel in both figures depicts the particular cases with 
the spontaneously long-range ordered ground state, whereas the lower panel shows the 
particular cases with the disordered spin liquid ground state. 
In accordance with the above statement, all temperature dependences of the heat capacity shown 
in both upper panels display a logarithmic singularity from the standard Ising universality 
class, which is associated with the continuous (second-order) phase transition between
the spontaneously ordered and disordered phases. Even although the specific heat curves 
displayed in Figs.~\ref{fig6} and \ref{fig7} have several common features, there are 
nevertheless a few important differences. In both cases, the marked round maximum is 
superimposed on the high-temperature tail of the specific heat singularity by considering 
a strong ferromagnetic intra-trimer interaction $J_{\rm H}/|J_{\rm I}| \gg 1$ (see Figs.~\ref{fig6}a and \ref{fig7}a). However, it can be also clearly seen that this round maximum is much more 
pronounced in the Ising model with $\Delta = 0$ than in the Ising-Heisenberg model with 
$\Delta = 1$. It is worthwhile to remark, moreover, that the round maximum in the 
high-temperature tail of the specific heat can be also observed when taking into account 
the antiferromagnetic intra-trimer interaction $J_{\rm H}/|J_{\rm I}| < 0$ (see Figs.~\ref{fig6}bc 
and \ref{fig7}bc), but in this particular case, the robust hump develops only if 
the ratio $J_{\rm H}/|J_{\rm I}|$ is close enough to the boundary value (\ref{eq14}). 
In this parameter space, the most important difference between the specific heat curves 
of the Ising and Ising-Heisenberg models lies in an appearance of the additional marked 
shoulder, which appears in the low-temperature tail of the heat capacity in the latter 
model only (Fig.~\ref{fig7}c), whereas this feature is obviously missing in the relevant 
dependence of the former model (Fig.~\ref{fig6}c). 

Last but not least, let us comment on temperature dependences of the specific heat by considering 
some typical cases that correspond to a spin system with the disordered spin liquid ground state 
(the lower panels in Figs.~\ref{fig6} and \ref{fig7}). Apparently, the temperature dependence 
with a single maximum is obtained by selecting the ratio $J_{\rm H}/|J_{\rm I}|$ exactly from 
the ground-state boundary (\ref{eq14}) between the spontaneously ordered and disordered phases 
(see Figs.~\ref{fig6}d and \ref{fig7}d). Note that this maximum is nevertheless much less symmetric 
for the Ising-Heisenberg model (Figs.~\ref{fig7}d) than for its Ising counterpart (Figs.~\ref{fig6}d), while both models exhibit a quite similar uprise of the additional low-temperature shoulder upon 
a consecutive slight decrease of the ratio $J_{\rm H}/|J_{\rm I}|$. As evidenced by Fig.~\ref{fig7}e, 
the specific heat of the Ising-Heisenberg model gradually looses its irregular shape upon 
further decrease of the interaction parameter $J_{\rm H}/|J_{\rm I}|$. It is noteworthy, moreover, 
that the specific heat curve with two marked rounded maxima appear by assuming a sufficiently 
strong antiferromagnetic intra-trimer interaction $J_{\rm H}/|J_{\rm I}| \ll -1$ (Figs.~\ref{fig6}f 
and \ref{fig7}f), whereas the more negative is the ratio $J_{\rm H}/|J_{\rm I}|$, the more evident 
is the double-peak structure of the zero-field specific heat curve. However, it should be also 
pointed out that the temperature dependence with the double-peak specific heat emerges in the Ising-Heisenberg model with $\Delta$ = 1 for much smaller geometric frustration (i.e. less 
negative ratio $J_{\rm H}/|J_{\rm I}|$) than for the Ising model with $\Delta$ = 0. 
Notwithstanding this observation, the specific heat curves with two well-separated maxima 
might be expected for the series of polymeric coordination compounds Cu$_9$X$_2$(cpa)$_6$ 
for which a rough estimate of the ratio $|J_{aa}/J_{ab}| \approx 40$ has been made according
to the experimentally measured susceptibility data.\cite{Mek98}

\section{\label{sec:conc}Concluding remarks}

The present article provides a survey of exact analytical results for the spin-1/2 Ising-Heisenberg 
model on the triangulated Kagom\'e (triangles-in-triangles) lattice, which has been proposed 
in order to shed light on a frustrated magnetism of the series of Cu$_9$X$_2$(cpa)$_6$ polymeric coordination compounds. The model under investigation has exactly been solved with the help 
of generalized star-triangle transformation that establishes a precise mapping relationship 
to the corresponding spin-1/2 Ising model on the Kagom\'e lattice. Exact results for the
ground-state and finite-temperature phase diagrams, as well as, several thermodynamic quantities 
such as the Helmholtz free energy, internal energy, entropy, and specific heat, have been derived 
and discussed in detail. 

Even though a theoretical description based on the hybrid Ising-Heisenberg model may not be fully realistic for true magnetic materials from the family of copper-based coordination compounds 
(Heisenberg model would be of course more adequate), it is quite reasonable to expect that the 
presented exact results illustrate many important features of the series of three isostructural compounds Cu$_9$X$_2$(cpa)$_6$. It is really plausible to argue that the suggested model correctly takes into account quantum effects closely connected with the stronger intra-trimer Heisenberg interaction $J_{\rm H}$, whereas the Ising character of the weaker monomer-trimer interaction $J_{\rm I}$ might be regarded as at least rather reasonable first-order approximation. 
This approximation should be altogether acceptable especially in the highly frustrated region 
$J_{\rm H}/|J_{\rm I}| \ll -2/(2 + \Delta)$, where the 'Ising' spins located at the monomeric 
sites are completely free to flip without any energy cost owing to a strong geometric frustration produced by the trimeric 'Heisenberg' spins. From this perspective, it is quite tempting to 
conjecture that the zero-field specific heat of Cu$_9$X$_2$(cpa)$_6$ compounds should exhibit 
a notable temperature dependence with two outstanding round maxima, which are well separated 
from each other because the antiferromagnetic intra-trimer interaction is roughly two orders 
of magnitude stronger than the ferromagnetic monomer-trimer interaction.

In agreement with our expectations, it also has been demonstrated that quantum fluctuations 
introduced through the non-zero exchange anisotropy $\Delta$ help to destroy spontaneous (ferromagnetic or ferrimagnetic) long-range ordering. As a matter of fact, it can be readily understood from Eq.~(\ref{eq14}) that the increase in the anisotropy parameter $\Delta$ suppresses a strength 
of the antiferromagnetic intra-trimer interaction that is needed to prevent spontaneous ordering. 
However, the most surprising finding to emerge from the present study closely relates to a substantial decline of the residual entropy of disordered spin liquid state, which appears on assumption that 
there is arbitrary but non-zero exchange anisotropy $\Delta$. It actually turns out that the 
quantum fluctuations partially lift a rather high macroscopic degeneracy of the disordered spin 
liquid state and consequently, the zero-point entropy of the quantum Ising-Heisenberg model 
$S_0/N_{\rm T} k_{\rm B} = 0.2806$ is for any $\Delta \neq 0$ significantly lower than 
the zero-point entropy of the semi-classical Ising model $S_0/N_{\rm T} k_{\rm B} = 0.4752$ 
achieved in the limiting case $\Delta = 0$. From this point of view, another interesting question 
arises whether or not the residual entropy of the spin-1/2 Heisenberg model on the triangulated 
Kagom\'e lattice will be completely removed by the order-from-disorder mechanism\cite{Vil80} 
when accounting for the Heisenberg character of the monomer-trimer interaction as well. 
In order to clarify this unresolved issue, it is necessary to perform a rather 
extensive exact numerical diagonalization of the full Heisenberg Hamiltonian, or, 
alternatively, it would be also possible to use the exact solution of the 
Ising-Heisenberg model as a starting point of more general perturbative treatment. 

\begin{center}
\textbf{Note added}
\end{center}

Shortly after our manuscript has been submitted for publication, Yao \textit{et al}. \cite{Yao08}
has completed a similar work dealing with the spin-1/2 Ising-Heisenberg model on the triangulated Kagom\'e lattice. The main difference between our procedure and the one of Yao \textit{et al}. \cite{Yao08} lies in the way of calculating the energy spectrum of Heisenberg trimer and its 
three enclosing Ising spins. In our procedure, we have firstly performed an exact diagonalization 
of the Heisenberg trimer in some generally non-uniform local field produced by three surrounding 
Ising spins and then, we have considered two symmetry-distinct configuration of the enclosing Ising spins in order to establish a precise mapping relationship with the corresponding spin-1/2 Ising model. 
In the procedure developed by Yao \textit{et al}.,\cite{Yao08} the authors first consider two symmetry-distinct configurations of the enclosing Ising spins and then, they diagonalize the 
Heisenberg trimer for both symmetry-distinct configurations of the surrounding Ising spins. 
In a such way, they were able to avoid rather cumbersome expressions for the roots of cubic 
equations, which otherwise occur when diagonalizing the Heisenberg trimer in some generally 
non-uniform local magnetic field. Despite this difference, both procedures turn out to be completely 
equivalent as it could be easily checked from a comparison of the effective mapping parameter 
$\beta J^{\rm eff}_{\rm kag}$ calculated from our Eqs.~(8-13) and Eqs.~(7-8,13-14) of Yao \textit{et al}.,\cite{Yao08} respectively. Exact results presented in both manuscripts for the ground-state 
and finite-temperature phase diagrams, Helmholtz free energy, internal energy and entropy 
are consequently in a very good accordance.

\begin{acknowledgments}
J. Stre\v{c}ka would like to thank Japan Society for the Promotion of Science for the award of postdoctoral fellowship (ID No. PE07031) under which part of this work was carried out. 
This work was supported by the Slovak Research and Development Agency under the contract 
LPP-0107-06. The financial support provided by Ministry of Education of Slovak Republic 
under the grant No.~VEGA~1/0128/08 is also gratefully acknowledged.
\end{acknowledgments}

\end{document}